\documentstyle[12pt]{article}
\begin{document}
\setlength{\unitlength}{1mm}

{\hfill Preprint E2-94-484}\\

{\hfill December 1994}\\

\begin{center}
{\Large\bf Temperature and Entropy of a Quantum Black Hole}
\end{center}
\begin{center}
{\Large\bf and Conformal Anomaly}
\end{center}

\bigskip\bigskip

\begin{center}
{\bf Dmitri V. Fursaev}\footnote{e-mail: fursaev@theor.jinrc.dubna.su}
\end{center}

\begin{center}
{{\it Bogoliubov Laboratory of Theoretical Physics, \\
Joint Institute for
Nuclear Research, \\
141 980, Dubna, Moscow Region, Russia}}
\end{center}

\bigskip\bigskip\bigskip

\begin{abstract}
Attention is paid to the fact that temperature of a classical
black hole can be derived from the extremality condition of its
free energy with respect to variation of the mass of a hole.
For a quantum Schwarzschild black hole evaporating massless
particles the same condition is shown to result in the following
one-loop temperature
$T=(8\pi M)^{-1} \left(1+\sigma (8\pi M^2)^{-1}\right)$
and entropy $S = 4\pi M^2 - \sigma\log M$
expressed in terms of the effective mass $M$ of a hole together with
its radiation and the integral of the conformal anomaly
$\sigma$ that depends on the field species.
Thus, in the given case quantum corrections to $T$ and $S$ turn
out to be completely provided by the anomaly. When it is
absent ($\sigma=0$), which happens in a number of
supersymmetric models, the one-loop expressions of $T$ and $S$ preserve the
classical form. On the other hand, if the anomaly is negative ($\sigma<0$)
an evaporating quantum hole seems to cease to heat up when
its mass reaches the Planck scales.
\end{abstract}

\vspace{7cm}

{\it PACS number(s): 04.60.+n, 12.25.+e, 97.60.Lf, 11.10.Gh}

\newpage
\baselineskip=.8cm

Black hole thermodynamics is known to possess a number of puzzles
like the meaning of black hole entropy, the information loss
problem and the operation of the generalized second law \cite{Bekenstein}.
The principal
difficulty on the way to their resolution is the lack of a consistent theory
of quantum gravity. Even investigation of quantum effects on the classical
curved backgrounds sometime represents a technical problem where
results can be obtained only approximately. This is a reason why
exactly solvable
two-dimensional models of black holes are of great interest at the
present moment \cite{2dmodels}.

The aim of this paper is to show how the one-loop corrections to the
temperature and entropy of the 4-dimensional
Schwarzschild black hole with massless quantum fields can be derived
explicitly  in a simple thermodynamical treatment based on the scaling
properties of the theory.

\bigskip\bigskip

To begin with, we remind that the energy $E$ and entropy $S$ of a canonical
ensemble at the temperature $\beta^{-1}$ can be derived from the free
energy $F(\beta)$ as follows:
\begin{equation}
E={\partial \over \partial \beta}(\beta F)~~~,~~~
S=\beta(E-F)~~~.
\label{1}
\end{equation}
These quantities for a system being at the fixed temperature change until
a system reaches a thermal equilibrium characterized by a minimum of $F$
\cite{Landau}.
In this state the condition of extremum for $F$
\begin{equation}
(\delta F)_{\beta}=0
\label{2}
\end{equation}
gives a relation between $\beta$ and other parameters of the ensemble.
Moreover, the first law of thermodynamics in its simplest form
\begin{equation}
\beta ^{-1}\delta S=\delta E
\label{3}
\end{equation}
turns out to be a consequence of (\ref{1}) and (\ref{2}).

Now, returning to thermodynamics of black holes, an extremality condition
of $F$, similar to (\ref{2}), can be used to relate the temperature of the
hole with its other parameters (mass, charge, etc.). To see this, we
make use of the Gibbons-Hawking approach to gravitational thermodynamics
\cite{Hawking}. In its
framework the free energy in the semiclassical approximation is given by the
Euclidean Einstein-Hilbert action $W_{cl}$ with suitably subtracted boundary
terms
\footnote{For simplicity we use the system of units $\hbar=c=G=k_B=1$.}
\begin{equation}
\beta F(\beta) = W_{cl}(\beta)=-{1 \over 16 \pi}\left(\int R \sqrt{g}d^4x
+2\int(K-K_0)\sqrt{h}d^3 x\right)~~~.
\label{4}
\end{equation}
This functional is taken on the corresponding gravitational instanton.
To elucidate the idea, consider as an example the Schwarzschild black hole
with the mass $m$. The Euclidean metric reads
\begin{equation}
ds^2=\left(1-{2m \over r}\right)d\tau ^2+\left(1-{2m \over r}\right)^{-1}dr^2
+r^2d\Omega^2
\label{5}
\end{equation}
and the presence of the temperature $\beta ^{-1}$ implies the periodicity
of this solution in $\tau$
\begin{equation}
0\le\tau\le \beta~~~.
\end{equation}
Although at arbitrary $\beta$ and $m$ the space (\ref{5}) has a conical
singularity at the horizon $r=2m$, the integral curvature in (\ref{4})
on such a space
is well-defined and it is non-zero. One can show \cite{cones}, \cite{FS1}
that on (\ref{5})
\begin{equation}
\int R\sqrt{g}d^4x=4\pi\left(1-{\beta \over 8\pi m}\right) A
\label{6}
\end{equation}
where $A=16\pi m^2$ is the area of the horizon. Plugging (\ref{6}) in
(\ref{4}) and taking into account the boundary terms, we get the free energy
\begin{equation}
F(\beta,m)=m-4\pi m^2\beta^{-1}~~~.
\label{7}
\end{equation}
The definitions (\ref{1}) applied to (\ref{7}) show that the energy of the
system equals the mass of the black hole, whereas its entropy is given
by the Bekenstein-Hawking formula
\begin{equation}
E=m~~~,~~~S=\frac 14 A~~~.
\label{8}
\end{equation}
Finally, finding the extremum of (\ref{7})  at fixed $\beta$
\begin{equation}
{\partial F(\beta,m) \over \partial m} = 1-8\pi m\beta ^{-1}=0
\label{9}
\end{equation}
one can gets the desired relation $\beta^{-1}=(8\pi m)^{-1}$ between
the temperature of the Hawking radiation and the mass of the hole.
However, as distinct from a normal canonical ensemble, a black hole
is the maximum of $F(\beta,m)$ rather than the minimum, which indicates
its well-known instability due to evaporation.

For simplicity we deal with the Schwarzschild black holes but one can show
that an analogous consideration for charged holes or those in a cavity of a
finite size is also possible. In particular, in these cases the value
of the Hawking temperature can also be obtained in the same manner from the
extremum of $F$ with respect to variation of the mass of the hole
when other parameters are fixed.
This fact is not
surprising. Indeed, even if the gravitational action (\ref{4}) is considered
on a class of manifolds admitting conical singularities, its extrema do not
change and they correspond to the smooth geometries \cite{cones}, \cite{FS1}.
A physical reason for
this is the absence of such a matter distribution over the horizon which
could give rise to conical defects.
On the other hand, different masses $m$ under fixed
$\beta$ are equivalent to Euclidean black holes with different
ranging of the time coordinate $\tau$, and their free energy has
the extremum when the conical singularity
vanishes, which is usually associated with the Hawking temperature
\cite{Hawking}.

\bigskip\bigskip

Consider now the black-hole thermodynamics with the one-loop
quantum corrections. We will be interested in the Schwarzschild hole
evaporating massless particles, so far as in this case quantum
effects can be evaluated explicitly. In quantum theory the effective
action and the free energy read
\begin{equation}
\beta F(\beta) = W(\beta)=W_{cl}(\beta)+W_Q(\beta)~~~.
\label{10}
\end{equation}
Here $W_{cl}$ is the classical action (\ref{4}) and $W_Q$ is a one-loop
contribution to it from $N_0$ scalar fields and, possibly, from
other fields of the higher spins ($h.s.$), which depends on the model in
question,
\begin{equation}
W_Q={N_0 \over 2}\log\det \nabla_{\mu}\nabla^{\mu} +~h.s.
\label{11}
\end{equation}
computed on the background space (\ref{5}).

Several remarks concerning (\ref{10}) and (\ref{11}) are in order.
To get rid off the standard ultraviolet divergences in $W_Q$, one should
include in (\ref{10}) higher order curvature terms. This also enables one
to remove completely \cite{FS2} the additional divergences in the
entropy of a black hole that are concentrated on its horizon \cite{'t Hooft}.
However, for the case of the Schwarzschild hole the role of $R^2$-terms
in the Lagrangian is reduced to irrelevant constant in the entropy
\cite{FS1}, and for this reason we omit these terms.
For massless fields there is also an infrared divergence in (\ref{11}).
It can be eliminated in the same manner \cite{Hawking} as for the classical
Einstein action (\ref{4}) by subtracting from the effective
gravitational functional $W$ (\ref{10}) additional terms given on a
distant spatial
boundary $r=r_0$. After the subtraction, the action (\ref{10}) turns out
be finite on (\ref{5})
and includes terms of the order $O(r_0^{-1})$ that
can be neglected in the limit $r_0\rightarrow\infty$.
We imply that infrared divergences are removed in such a way but do not
write the boundary terms in (\ref{10}) explicitly since their form is
also irrelevant for further consideration.

For the functional (\ref{10}) taken on the space (\ref{5})
the only free parameter, apart from $\beta$, is the mass $m$
of the hole and as in classical theory we can consider its variation with
respect to this parameter. Thus, the extremality
condition of $F(\beta)$ can be represented as
\begin{equation}
\beta - 8\pi m + \left({\partial W_Q \over \partial m}\right)_{\beta}=0~~~.
\label{12}
\end{equation}
Equation (\ref{12}) indicates a correction $\partial _m W_Q$
to the Hawking temperature
which can be calculated as follows. Consider the scaling properties of
$W_Q$ that depends on $m$ through the background
metric (\ref{5}) and on $\beta$ through the boundary
conditions. Assuming $W_Q$ to be a renormalized action, one can write
$$
W_Q\left(\beta, g_{\mu\nu}(m)\right)=W_Q\left(\beta\alpha^{-1},\alpha^2
g_{\mu\nu}(m\alpha^{-1})\right)=
$$
\begin{equation}
W_Q\left(\beta\alpha^{-1},g_{\mu\nu}(m\alpha^{-1})\right)+
\left(\int T_{\mu}^{\mu}\sqrt{g}d^4x - a_{surf}(\beta
\beta_H^{-1})\right)\log\alpha
\label{13}
\end{equation}
where $\alpha$ is an arbitrary parameter and $\beta_H\equiv8\pi m$.
The last term in the r. h. s. of (\ref{13}) appears
due to the breaking in $W_Q$ of the conformal invariance to be held for
classical massless fields. It includes
the standard trace anomaly of the renormalized stress tensor
$T_{\mu}^{\mu}=-(16\pi^2)^{-1}a_2$ determined by
the $a_2$-coefficient in the DeWitt-Schwinger proper time expansion
\cite{a16}. In our case, it is
\begin{equation}
\int_0^{\beta}d\tau\int_{2m}^{\infty}r^2dr\int d\Omega~T_{\mu}^{\mu}
=\sigma {\beta \over \beta_H}
\label{14}
\end{equation}
where $\sigma$ is the integral of the trace anomaly at $\beta=\beta_H$ that
depends on the numbers $N_s$ of the fields with the spin $s$
entering in the model \cite{a16}, \cite{Duff}
\begin{equation}
\sigma ={1 \over 45}\left(-N_0-\frac 74 N_{1/2}+13 N_1 +{233 \over 4}N_{3/2}
-212N_2\right)~~~.
\label{15}
\end{equation}
There is also an additional anomalous term $a_{surf}(\beta \beta_H^{-1})$
in the transformation law of $W_{Q}$ due to the conical singularities of the
background manifold;
$a_{surf}(\beta \beta_H^{-1})$ is an integral over the horizon surface which
has been exactly found for the scalar determinants in \cite{Dowker}.
However, the only thing
important for us is that this addition disappears at the Hawking temperature
\begin{equation}
a_{surf}(\beta \beta_H^{-1})=0~~~,~~~\beta=\beta_H=8\pi m~~~.
\label{16}
\end{equation}
It is suitable to choose $\alpha=m$ and represent (\ref{13}) as
\begin{equation}
W_Q\left(\beta, g_{\mu\nu}(m)\right)=
W_Q\left(\beta m^{-1}, g_{\mu\nu}(m=1)\right) + \left(\sigma\beta
\beta_H^{-1}
-a_{surf}(\beta \beta_H^{-1})\right)\log m\equiv f(\beta m^{-1}, m)~.
\label{17}
\end{equation}
This immediately results in the relation
$$
\left({\partial W_Q \over \partial m}\right)_{\beta}=
\left({\partial f \over \partial m}\right)_{(\beta m^{-1})}
-{\beta \over m}\left({\partial f \over \partial \beta}\right)_{m}=
$$
\begin{equation}
\frac 1m\left(\sigma {\beta \over \beta_H} - a_{surf}\left(\beta \beta_H^{-1}
\right)
\right)
-{\beta \over m}\left({\partial W_Q \over \partial \beta}\right)_{m}~~~.
\label{18}
\end{equation}
Inserting (\ref{18}) into condition (\ref{12}) we have
\begin{equation}
\beta -8\pi m-{\beta \over m}\left(\left({\partial W_Q \over \partial \beta}
\right)_m - {\sigma \over \beta_H}\right)-\frac 1m a_{surf}\left(\beta
\beta_H^{-1}\right)=0~~~
\label{19}
\end{equation}
and $\beta $ can be found from (\ref{19})
by iteration in the Planck constant $\hbar$ as a series. Thus, taking into
account
(\ref{16}), one obtains the expression
\begin{equation}
\beta =8\pi\left(m+\left({\partial W_Q \over \partial \beta}
\right)_m - {\sigma \over 8\pi m}\right)+ O(\hbar ^2)~~~.
\label{20}
\end{equation}
The quantity $\partial _{\beta} W_Q$ in (\ref{20}) is the thermal energy
of quantum fields associated with the radiation of a hole and it is
an unknown functional of the background metric. Fortunately, there is no need
to calculate it explicitly so far as
equation (\ref{20}) can be rewritten through the total internal
energy of the system
\begin{equation}
E={\partial \over \partial \beta}(\beta F)_m = m + \left({\partial W_Q
\over \partial \beta}\right)_m\equiv M~~~.
\label{21}
\end{equation}
The constant $M$ can be considered as
the effective gravitational mass including the energy of the radiation
and, as distinct from the classical mass $m$, it is an observable parameter
of the theory. In terms of $M$ and in the first order in $\hbar$, $\beta$
takes the simple form
\begin{equation}
\beta = 8\pi\left(M - {\sigma \over 8\pi M}\right)
\label{22}
\end{equation}
(replacing $m$ by $M$ in the anomalous term in (\ref{20}) results in
a correction $O(\hbar ^2)$). Consequently, the one-loop
temperature reads
\begin{equation}
T=T_H(M)\left(1+{\sigma \over 8\pi M^2}\right)
\label{23}
\end{equation}
where $T_H(M)=(8\pi M)^{-1}$ is the classical Hawking temperature defined
for a hole with the mass $M$. The one-loop entropy can be recovered from
(\ref{23}) by making use of Clausius's rule
\begin{equation}
S=\int {dM \over T}= 4\pi M^2 - \sigma\log M
\label{24}
\end{equation}
and it differs from the Bekenstein-Hawking entropy by the logarithmic term.
Another way to derive (\ref{24}) is to use the statistical-mechanical
definition (\ref{1}) of $S$. Equations (\ref{23}) and (\ref{24}) represent
the main result of this paper.
Remarkably that $T$ and $S$ can be found explicitly and coming out is the
only new coefficient $\sigma$ of the field species given by
equation (\ref{15}).

The temperature $T$ has been derived from the extremum of the one-loop free
energy or, which is the same, of the effective gravitational action $W$,
see (\ref{10}). Although $W$ is a non-trivial functional of the metric,
one can expect that it possesses the same
property as the Euclidean Einstein action (\ref{4}) when quantum effects are
weak and has the extrema on
the smooth manifolds with the black hole geometry similar to (\ref{5}).
This seems to be a natural assumption because, as was pointed out,
non-smooth solutions with conical singularities would correspond to
some specific matter distribution concentrated on the horizon surface of
a hole. Therefore, in quantum case one can repeat the same arguments
given above for the classical action (\ref{4}) and relate the extremum
(\ref{12})
of $W$ with vanishing of the conical singularity
for the Schwarzschild solution deformed by one-loop quantum corrections.
After that
the temperature (\ref{23}) should be related with the one-loop surface gravity
$k$ as $T=(2\pi)^{-1}k$ and, hence,
one can identify it with the temperature of the
Hawking radiation in presence of the back reaction.

Let us discuss these results. As is seen from (\ref{23}) and (\ref{24}),
in the model in question the difference of $T$ and $S$ from their classical
form is completely provided by the conformal anomaly (\ref{15}).
In this context it is interesting to pay attention to
the role played by the anomalous
trace in two dimensional theory where it determines the flux of the
Hawking radiation \cite{CF}. In four dimensions
the anomaly is known to be absent in the models of $N=8$ and $N=4$ supergravity
and in the $N=4$ super Yang-Mills theory \cite{Duff}.
Thus, following from (\ref{23}) and (\ref{24}) is an
interesting consequence that for these models the one-loop corrections
can change the mass of the Schwarzschild black hole, whereas the form
of the thermodynamics is left the same as in the classical case.
In general, the behavior of $T$ and $S$ depends on the sign of $\sigma$.
The latter is positive when spins 1 and 3/2 dominate in the theory
and then quantum effects accelerate evaporation of the hole by increasing
its temperature.

A qualitative difference from the classical black hole thermodynamics
appears for the negative anomaly $\sigma<0$, when the scalars,
fermions and gravitons prevail.
Then, the increase in the hole temperature $T$ slows down. Moreover,
in this case,
when mass approaches the Planck scales $M\simeq \sqrt{\sigma}M_{Planck}$,
$T$ reaches a maximum and after that starts to decrease, a hole cools down.
Surely, in this domain the one-loop approximation is not reliable and
another, probably, nonperturbative treatment is needed. However, if (\ref{23})
is used for extrapolation to the Planck region, it shows that temperature
is null for some small or zero values of $M$, which can be interpreted as the
end of evaporation. If this were actually true, the black hole evaporation
would finish by a pure vacuum state. This eventually would remove the
information
loss paradox \cite{Bekenstein}.

Our analysis would be incomplete without comparing equations (\ref{23}) and
(\ref{24}) with the one-loop quantities derived by taking directly
into account the back reaction caused to the Schwarzschild metric
by the quantum matter \cite{York}, \cite{Lousto}.
However, to employ the back-reaction method, one
needs the renormalized stress tensor that is known for the
Schwarzschild hole in 4-dimensions only in the Page approximation \cite{Page}.
Nevertheless, there is a qualitative agreement between equation (\ref{23}) and
that reported in \cite{York}, \cite{Lousto}. In particular, the maximum of
the radiation temperature was also mentioned in \cite{Lousto} for the
gravitation dominated matter.
It is also worth pointing out that a logarithmic dependence of the one-loop
black hole entropy on the mass, similar to (\ref{24}), has been
found out in a number of two-dimensional models, for instance in
\cite{Fiola}, and has been argued to occur in the membrane approach to the
description of black holes \cite{Maggiore}.

One should remark in conclusion that the reason why the simple expressions
for $T$ and $S$ have been obtained in our method is that the Schwarzschild
metric possesses the only dimensional parameter $m$. Thus, it is interesting
to repeat the analysis for more general black hole geometries and massive
quantum fields and see how the properties of the considered model can
change.

\bigskip\bigskip

The author is very grateful to Professor Igor Novikov for hospitality at
Theoretical Astrophysics Center in Copenhagen and helpful discussions
and also thanks Sergey Solodukhin for valuable contacts.
This work was partially supported by Russian Foundation for Fundamental
Science, grant N 94-02-03665-a.


\begin{thebibliography} \\

\bibitem{Bekenstein} J.D. Bekenstein, {\it Do we understand black hole
entropy?}, gr-qc/9409015.
\bibitem{2dmodels} C.G. Callan, S.B. Giddings, J.A. Harvey and A. Strominger,
Phys. Rev. {\bf D45}, R1005 (1992); J.G. Russo, L. Susskind and L. Thorlacius,
Phys. Rev. {\bf D46}, 3444 (1992), Phys. Rev. {\bf D47}, 533 (1993).
\bibitem{Landau} L.D. Landau and E.M. Lifshitz, {\it Statistical Physics}
(Pergamon Press, London - Paris, 1958).
\bibitem{Hawking} G.W. Gibbons, S.W. Hawking, Phys. Rev. {\bf D15}, 2752
(1977); S.W. Hawking in General Relativity, ed. by S.W. Hawking and W. Israel
(Cambridge Univ. Press, Cambridge, 1979).
\bibitem{cones} G.H. Hayward and J. Louko, Phys. Rev. {\bf D42}, 4032
(1990).
\bibitem{FS1} D.V. Fursaev and S.N. Solodukhin, {\it On description of the
Riemannian geometry in presence of conical singularities}, in preparation.
\bibitem{FS2} D.V. Fursaev and S.N. Solodukhin, {\it On one-loop
renormalization of black hole entropy}, preprint E2-94-462, hep-th/9412020.
\bibitem{'t Hooft} G.'t Hooft, Nucl. Phys. {\bf B256}, 727 (1985).
\bibitem{a16} N.D.Birrell and P.C.W.Davies, {\it Quantum Fields in Curved
Space} (Cambridge Univ. Press, New York 1982).
\bibitem{Duff} S.M. Christensen and M.J. Duff, Phys. Lett. {\bf B76}, 571
(1978); M.J. Duff, Class. Quantum Grav. {\bf 11}, 1387 (1994).
\bibitem{Dowker} J.S. Dowker, {\it Effective actions with fixed points},
preprint MUTP/94/14, hep-th/9406144.
\bibitem{CF} S.M. Christensen and S.A. Fulling, Phys. Rev. {\bf D15}, 2088
(1977).
\bibitem{York} J.W. York, Phys. Rev. {\bf D31}, 775 (1985).
\bibitem{Lousto} C.O. Lousto and N. Sanchez, Phys. Lett. {\bf 212},
411 (1988).
\bibitem{Page} D.N. Page, Phys. Rev. {\bf D25}, 1499 (1982).
\bibitem{Fiola} T.M. Fiola, J. Preskill, A. Strominger and S.P. Trivedy,
{\it Black hole thermodynamics and information loss in two dimensions},
preprint CALT-68-1918, hep-th/9403137; S.N. Solodukhin, {\it The conical
singularity and
quantum corrections to entropy of black hole}, preprint JINR E2-94-246,
Phys. Rev. D to be published.
\bibitem{Maggiore} C.O. Lousto and M. Maggiore, {\it On the energy spectrum
of quantum black holes}, preprint IFUP-TH 53/94, gr-qc/9411037.
\end{thebibliography}
\end{document}